\documentclass[seceq]{ptptex}

\usepackage{amsmath,amsfonts,latexsym,amssymb}
\usepackage{verbatim,amsthm,graphicx}
\usepackage{wrapft}

\def\FigDir{.}

\usepackage{amsmath}
\usepackage{amssymb}
\usepackage{mathrsfs}

\def\defscript{\mathscr}

\def\LL{{\defscript L}}

\def\ifempty#1{\def\tmpdata{#1}\ifx\tmpdata\empty }

\def\linebreak{\hfill\break}

\def\bra<#1|{\langle #1\rvert}
\def\ket|#1>{\lvert#1 \rangle}
\def\braket<#1|#2>{\langle #1|#2 \rangle}

\def\pfrac#1#2{\left(\frac{#1}{#2}\right)}
\def\const{\text{const}}
\def\otop#1{\hbox{$#1\kern-0.1em$\llap{\hbox{\raise1.7ex\hbox{$\scriptstyle\circ$}}}} }

\def\inpare#1{\left(#1\right)}
\def\bigpare(#1){\left(#1\right)}
\def\inrbra#1{\left\{ #1 \right\}}
\def\insbra#1{\left[ #1 \right]}

\def\bigbra[#1]{\left[ #1 \right]}

\def\t{\tilde }
\def\d{\dot }

\def\b{\bar }

\def\therefore{\mbox{\setbox0=\hbox{X}\hbox{$\ldotp$}\raise0.7\ht0\hbox{$\ldotp$}\hbox{$\ldotp$}} \quad }
\def\because{\mbox{\setbox0=\hbox{X}\raise0.7\ht0\hbox{$\ldotp$}\hbox{$\ldotp$}\raise0.7\ht0\hbox{$\ldotp$}}\kern0pt }

\def\bm#1{\boldsymbol{#1}}

\def\ZR{{{\mathbb Z}}}

\def\CF{{{\mathbb C}}}

\def\UG{{\rm U}}

\def\SU{{\rm SU}}

\def\upin{\hbox{\setbox0=\hbox{$\cup$} \vrule width 0.05 \wd0 height \ht0 depth 0pt \kern - 0.5\wd0 \box0 }}
\def\Frac(#1/#2){\left(\frac{#1}{#2}\right)}

\def\Im{{\rm Im\,}}
\def\Re{{\rm Re\,}}

\def\sdprod{\mathrel{{\setbox0=\hbox{$\displaystyle\times$}\lower0.3\wd0\hbox{$\stackrel{\box0}{\scriptstyle\sim}$}}}}

\def\w{\wedge}

\def\tosigma#1,{%
    \ifx\tmpindex\relax \def\tmpindex{#1} \let\next=\tosigma
    \else \ifnum\tmpindex=0 1 \else \sigma_\tmpindex \fi
          \ifx#1\relax  \let\next=\relax
          \else \otimes \let\next=\tosigma \def\tmpindex{#1} \fi
    \fi \next}
\def\tspb(#1){\let\tmpindex=\relax\tosigma#1,\relax,}
\def\Order#1{{\rm O}\!\left(#1\right)}

\def\pd{\partial}

\def\CP{{\CF P}}

\def\Eq#1{\begin{equation} #1 \end{equation}}

\def\Eqr#1{\begin{eqnarray} #1 \end{eqnarray}}
\def\Eqrn#1{\begin{eqnarray*} #1 \end{eqnarray*}}
\def\Eqrsub#1{\begin{subequations}\Eqr{#1}\end{subequations}}
\def\Eqrsubl#1#2{\begin{subequations}
  \expandafter\ifx\csname Rlabel\endcsname \relax \label{#1}
  \else \Rlabel{#1} \fi \Eqr{#2}\end{subequations}}
\def\Bitm{\begin{itemize}}
\def\Eitm{\end{itemize}}
\def\Blist#1#2{\begin{list}{#1}{\parsep=0pt \itemsep=0pt%
  \listparindent=0pt #2}}
\def\Elist{\end{list}}
\long\def\ignore#1#2{\def\ignoreflag{#1}\long\def\tmptext{#2}
  \ifnum\ignoreflag>1 #2 \fi}

\begin{document}
\title{\large Repulsons in the Myers-Perry Family}

\author{%
Gary {\sc Gibbons}$^{a,}$%
\footnote{E-mail: gwg1@amtp.cam.ac.uk}
and
Hideo {\sc Kodama}$^{b,}$%
\footnote{E-mail: hideo.kodama@kek.jp}
}
\inst{${}^a$D.A.M.T.P., Cambridge University,\\
Wilberforce Road, Cambridge CB3 0WA, U.K.\\
${}^b$IPNS, KEK and the Graduate University of Advanced Studies,\\
1-1 Oho, Tsukuba 305-0801, Japan}

\abst{
In this paper, we show that curvature-regular asymptotically flat solitons with negative mass are contained in the Myers-Perry family of odd spacetime dimensions. These solitons do not have a horizon, but instead a conical singularity of quasi-regular nature surrounded by naked CTCs. This quasi-regular singularity can be made regular for a set of discrete values of angular momentum, at least for the Myers-Perry solutions with $\UG(N)$ symmetry. Although the time coordinate is required to have a periodicity at infinity and the spatial infinity becomes a lens space $S^{D-2}/\ZR_n$, the corresponding spacetime is simply connected, and the CTCs cannot be eliminated by taking a covering spacetime.
}

\maketitle

\section{Introduction}

In four dimensions, although no general theorem forbids an asymptotically flat regular simply-connected solution with naked CTCs,  no such solution has been found. For example, 
the Kerr solution has CTCs only when it has a negative mass  and  naked ring singularity. The
Kerr-Newman solution has  naked CTCs even for positive mass  but only in the superextreme case with naked ring singularity. Similarly, the Tomimatsu-Sato solution with $\delta=2$\cite{Tomimatsu.A&Sato1972} has double horizons and circles generated by the rotational Killing vector become time-like near the $z$-axis segment connecting two horizons, but the spacetime has the well-know naked ring singularity and a conical singularity along the segment\cite{Kodama.H&Hikida2003}. Here, the asymptotically flatness is a crucial condition, because we have the Taub-NUT solution as a famous example with a curvature-regular vacuum solution with CTCs if we do not impose that condition. The existence of matter also tends to allow solutions with CTCs, as illustrated by the Kerr-Newman solution and the G\"odel solution.

When we go to higher dimensions, the situation changes. For example,  Gibbons and Herdeiro\cite{Gibbons.G&Herdeiro1999} examined the causal structure of the BMPV solution\cite{Breckenridge.J&&1997,Kallosh.R&Rajaraman&Wong1997}, which is a BPS solution with two parameters $\mu$ and $\omega$ to the five-dimensional gauged supergravity theory obtained from the type IIA supergravity by compactification over K3$\times S^1$:
\Eqrn{
&& ds^2=-\Delta^2(dt-\Omega \chi^3)^2 + \frac{dr^2}{\Delta^2}
 + r^2 ds^2(S^3),\\
&&\qquad \Delta=1-\frac{\mu}{r^2},\quad
     \Delta\Omega=\frac{\mu\omega}{r^2},\\
&& A=\frac{\nu}{r^2}( dt + \omega \chi^3),
}
where $\chi^3$ is a normalised invariant 1-form on $S^3$ (see \S\ref{sec:DiskBundle} in the text for its definition), and $A$ is an Abelian gauge field of the RR origin. They noticed that there always exist CTCs if $\omega \mu\neq0$ but confirmed that for the parameter range $\omega^2<\mu$, these CTCs are confined inside degenerate non-rotating horizon at $r^2=\mu$ where $\Delta=0$. Thus, the solution represents a regular rotating black hole with charge in five dimensions. However, they found that for $\omega^2>\mu$, these CTCs come out to the region $\Delta>0$ and that the locus $r^2=\mu$ is not a horizon anymore, because no geodesic can penetrate the "surface" $r^2=\mu$. Because of this property, they named the surface a repulson. Later, similar BPS solutions were found in other supergravity theories in five dimensions\cite{Cvetic.M&&2005A,Gauntlett.J&&2003}. 

Thus, the existence of CTCs, regularity and asymptotically flatness can coexist in higher dimensions. Although these are examples in supergravity theories, there is another example suggesting that such a situation can be realised even for a vacuum system. In fact, Myers and Perry pointed out in \citen{Myers.R&Perry1986} that when the total mass is negative and the spacetime dimension is odd, their solution can be extended regularly across the apparent singularity at $r=0$ to the region $r^2<0$ with CTCs up to the point where the volume of the Killing orbit vanishes. They regarded this solution as representing an asymptotically flat 'black hole with negative mass' and argued that its apparent contradiction with the positive energy theorem is evaded due to the pathological feature of the solution that causality is violated outside the 'horizon'.  

The main purpose of the present paper is to point out that the 'horizon' they found is not a horizon, but rather a quasi-regular singularity similar to the repulson when all the angular momentum parameters are equal. We will show further that the singularity disappears for a set of discrete values of the angular momentum after a periodic identification in time. Thus, we can construct asymptotically flat regular spacetimes with negative mass and naked CTCs. 

The paper is organised as follows. In the next section, as the simplest example, we study the internal structure of the Myers-Perry solution in five dimensions in detail and show that something strange happens when the mass becomes negative. In particular, we point out that the locus which is a horizon boundary for positive mass appears to turn to a regular NUT-type internal submanifold. In Section \ref{sec:DiskBundle}, as preliminaries for construction of repulsons in the following section, we explain the basic mathematical facts about the $S^1$ bundle over $\CP^{N-1}$ and associated disk bundle over $\CP^{N-1}$. Then, in Section \ref{sec:repulsons}, we show that the internal structure of the Myers-Perry solution with equal angular momentum parameters in odd spacetime dimensions exhibits similar behavior as that in the five-dimensional case, and construct regular soliton solutions from them with negative mass. The final section is devoted to a summary and discussion.

\section{Internal Structure of 5D Myers-Perry Solution}
\label{sec:5DMP}

In this section, we examine the internal structure of the MP solution in five dimensions as the simplest case and point out its peculiar feature in the negative mass case.

\subsection{Singularities}

In terms of the Boyer-Lindquist coordinates $(t,{\phi_1},{\phi_2},r,\theta)$, the 5D Myers-Perry solution can be written as\cite{Myers.R&Perry1986,Hawking.S&Hunter&Taylor-Robinson1999,Gibbons.G&&2005}
\Eqr{
ds^2 &=& \frac{r^2\rho^2}{\Delta} dr^2 + \rho^2d\theta^2
\notag\\
   && +(r^2+a^2)\sin^2\theta d{\phi_1}^2+ (r^2+b^2)\cos^2\theta d{\phi_2}^2
   \notag\\
   && -dt^2 + \frac{\mu }{\rho^2}\insbra{dt+a\sin^2\theta d{\phi_1} + b\cos^2\theta d{\phi_2}}^2,
\label{metric:5DMP:BL}
}
where
\Eq{
\Delta :=(r^2+a^2)(r^2+b^2)- \mu  r^2,\quad
\rho^2 := r^2 + a^2 \sin^2\theta + b^2 \cos^2\theta.
}
%
The determinant of the metric is given by
\Eq{
-g = r^2\rho^4 \sin^2\theta\cos^2\theta.
}
The Kretchman invariant of this metric is written
\Eq{
R_{\mu\nu\lambda\sigma}R^{\mu\nu\lambda\sigma}
=24 \mu ^2 \frac{(3r^2-a^2\cos^2\theta-b^2\sin^2\theta)(r^2-3a^2\cos^2\theta-3b^2\sin^2\theta)}{\rho^{12}}.
}
From this we find that the surface $\rho^2=0$ is scalar curvature singularity.

The metric in the Boyer-Lindquist coordinates is apparently singular at $r=0$ and at $\Delta=0$ in addition to  $\rho=0$, apart from the angular coordinate singularities at $\theta=0,\pi/2$. Among these, the points with $r=0$ are not real singularity as pointed out by Myers and Perry\cite{Myers.R&Perry1986}, because the metric can be expressed in terms of $x=r^2$ as
\Eqr{
ds^2 &=& \frac{\rho^2}{4\Delta} dx^2 + \rho^2d\theta^2
\notag\\
   && +(x+a^2)\sin^2\theta d{\phi_1}^2+ (x+b^2)\cos^2\theta d{\phi_2}^2
   \notag\\
   && -dt^2 + \frac{\mu }{\rho^2}\insbra{dt+a\sin^2\theta d{\phi_1} + b\cos^2\theta d{\phi_2}}^2.
\label{metric:5DMP:x}
}
Now, $\Delta$ and $\rho^2$ become polynomials in $x$ 
\Eq{
\Delta=(x+a^2)(x+b^2)-\mu x,\quad
\rho^2=x+a^2\cos^2\theta+b^2\sin^2\theta,
}
and $-g$ are given as
\Eq{
-g= \frac{1}{4}\rho^4 \sin^2\theta \cos^2\theta.
}
We will show later that $\Delta=0$ is horizon for $\mu>0$ and is not singularity in general. 

\subsection{Structure of Killing orbits}

\begin{figure}
\centerline{
\includegraphics*[height=8cm]{\FigDir/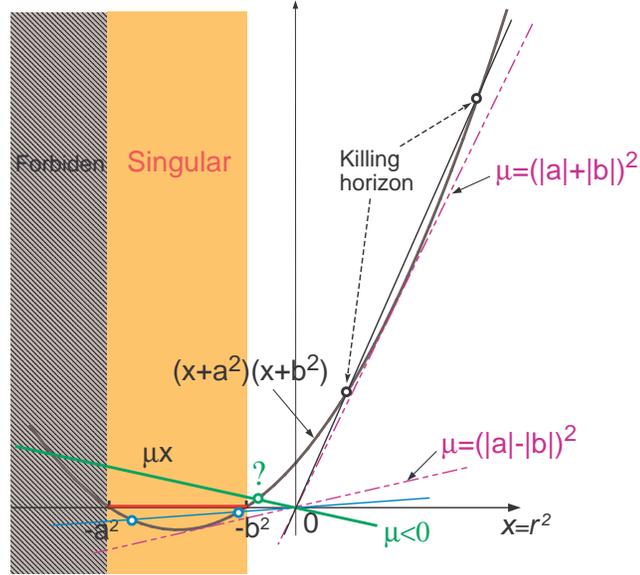}
}
\caption{\label{fig:xh} Horizon position $x_h$}
\end{figure}

On a Killing horizon, the induced metric on the orbit space of the Killing vectors $\xi=\pd_t, \eta_1=\pd_{\phi_1}$ and $\eta_2=\pd_{\phi_2}$ becomes degenerate. From
\Eq{
\left|g_{a b}\right|_{a,b=t,{\phi_2},{\phi_1}}=-\Delta \sin^2\theta\cos^2\theta,
}
this condition can be written 
\Eq{
\Delta\equiv (x+a^2)(x+b^2)-\mu x=0.
\label{horizoneq:x}
}

When $ab\neq0$, for 
\Eq{
\mu <(|a|-|b|)^2,
\quad
\text{or}\quad
\mu >(|a|+|b|)^2,
}
this equation has two roots:
\Eq{
x=\frac{1}{2}\inrbra{\mu -a^2-b^2\pm\insbra{(a^2+b^2-\mu )^2-4a^2b^2}^{1/2}}
.
}
In the case $\mu >(|a|+|b|)^2$, these roots are both positive, and the corresponding hypersurfaces are regular horizons, as we show in Appendix \ref{app:horizon}. In contrast, in the case $0<\mu <(|a|-|b|)^2$, both roots are smaller than $-b^2$ and a naked singularity appears because the spacetime becomes singular at points where $\rho^2$ vanishes. In the special case with $a^2>b^2=0$, the roots of $\Delta$ are given by
\Eq{
x=0, \mu -a^2.
}
Hence, the surface  $r=0$ is always a Killing horizon, but it contains a singularity. This singularity is hidden behind a horizon only when $\mu >a^2$.

Finally, in the case $\mu <0$, one of the roots, $x=x_h$, is larger than $-b^2$ and in the region with $\rho^2>0$, while the other is smaller than $-a^2$. This apparently indicates that even in the negative mass case, the spacetime has a regular horizon and a regular domain of outer communication\cite{Myers.R&Perry1986}. However, this is not the case. In fact, the spacetime has pathological features around $x=x_h$, when $x_h<0$. For example, the Killing vector $k$ whose norm vanishes at $x=x_h$ is given by
\Eq{
k=\pd_t - \frac{a}{x_h+a^2} \pd_{\phi_1} - \frac{b}{x_h+b^2}\pd_{\phi_2},
}
where the last term does not exist for $b=0$ and $x_h=0$. The norm of this vector is expressed as
\Eqr{
&& k\cdot k = \frac{(x-x_h)P}{\rho^2(x_h+a^2)^2(x_h+b^2)^2},\\
&& P = (a^2b^2-x_h^2)\rho_h^4+ \inrbra{(a^2-b^2)(a^2b^2-x_h^2)\cos^2\theta + a^2(x_h+b^2)^2}(x-x_h).
}
Hence, near the horizon, we have
\Eq{
k\cdot k = - \frac{x_h\Delta\rho^2}{(x_h+a^2)^2(x_h+b^2)^2}(1+\Order{\Delta}).
}
From this, it follows that for $x_h<0$, $k$ becomes spacelike outside the horizon. This behavior seems to be closely related to the fact that there appears a causality violating region in $x>x_h$ for $x_h<0$.  We will show in \S\ref{sec:repulsons} that in this case, $x=x_h$ is  not a Killing horizon but rather a quasi-regular singularity that is similar to the repulson first found by Gibbons and Herdeiro for supersymmetric rotating black holes in five dimensions\cite{Gibbons.G&Herdeiro1999,Cvetic.M&&2005A}.

\subsection{CTCs}

\begin{figure}
\centerline{
\includegraphics*[height=8cm]{\FigDir/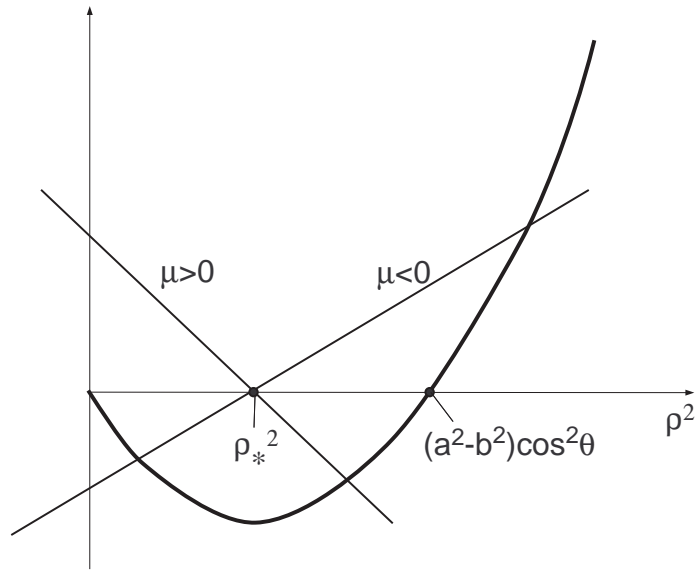}
}
\caption{\label{fig:D1}Causality violating region.  In the region where the thick curve is below the straight line, $\Delta_1$ becomes negative. In this figure, $\rho_*^2=(a^2-b^2)^2\sin^2\theta\cos^2\theta/(a^2\sin^2\theta+b^2\cos^2\theta)$.}
\end{figure}

In general, it is rather hard to see exactly whether there exists a CTC or not because it is a global problem. In the present case, however, it can be reduced to a simple local problem for the following reason. 

Let us suppose that there exists a causal closed curve $\gamma$. Then, since $g_{xx}>0$ in the region where $\Delta>0$ and the remaining part increases with $x$ in the metric \eqref{metric:5DMP:x}, the projection of $\gamma$ onto the $x=x_m=\const$ surface should be also a causal closed curve, where $x_m$ is the minimum value of $x$ on $\gamma$. Let us denote that curve by $\gamma'$. Then, it is a closed curve on a line bundle over $S^3$, where the fiber is generated by the Killing vector $\pd_t$, unless the time coordinate $t$ is not identified periodically. Thus, $t$ takes a maximum value at some point $p$ on $\gamma'$, where $dt$ vanishes for the tangent vector to $\gamma'$. Since $g_{\theta\theta}=\rho^2>0$, the projection of the tangent vector onto the 2-surface passing through $p$ generated by the two Killing vectors $\pd_{\phi_1}$ and $\pd_{\phi_2}$ should be causal. If the original curve $\gamma$ is a CTC, this vector obtained by projection should be also timelike. Conversely, if the metric of the 2-torus generated by $\pd_{\phi_1}$ and $\pd_{\phi_2}$ contains a time-like direction, we can always find a CTC on this surface, because a geodesic which is not closed on a 2-torus is ergodic. 
Therefore, there exists a CTC in our spacetime in the region $x>x_h$ if and only if the symmetric matrix $\Phi$ of degree 2 defined by
\Eq{
\Phi:= \inpare{g_{ij}}_{i,j={\phi_1}.{\phi_2}},
}
has a negative eigenvalue. 

The determinant of this matrix is given by
\Eq{
\Delta_1 := \det \Phi= \frac{D_1}{\rho^2}\sin^2\theta\cos^2\theta,
}
where
\Eqr{
 D_1& =& \inrbra{a^2b^2+(a^2\sin^2\theta+b^2\cos^2\theta)x}\mu  \notag\\
    && + x^3+\inrbra{a^2(1+\cos^2\theta)+b^2(1+\sin^2\theta)}x^2 \notag\\
    &&+(2a^2b^2+a^4\cos^2\theta+b^4\sin^2\theta)x
      +a^2b^2(a^2\cos^2\theta+b^2\sin^2\theta) \notag\\ 
  &=&\rho^2\inpare{\rho^2-(a^2-b^2)\cos^2\theta}\inpare{\rho^2+(a^2-b^2)\sin^2\theta}
  \notag\\
  && + \mu \inrbra{(a^2\sin^2\theta+b^2\cos^2\theta)\rho^2-(a^2-b^2)^2\sin^2\theta\cos^2\theta}.
}
Hence, the sign of $\Delta_1$ is determined by the difference of a fixed function cubic in $\rho^2$ and a function that is linear in $\rho^2$ and proportional to $\mu $, as shown in Fig. \ref{fig:D1}. For definiteness, we can assume that $a^2\ge b^2$ without loss of generality. Then, from this figure, we see that $D_1$ always becomes negative for some positive values of $\rho^2$ and $\theta\neq\pi/2$ except for $a^2=b^2$, for which $D_1$ changes its sign only when $\mu <0$.  However, because $D_1$ is non-negative in $x>0$ for $\mu >0$, there occurs no causality violation at or outside the outer horizon for $\mu >(|a|+|b|)^2$.

In contrast, for $\mu <0$, there appears a causality violating region around $x=x_h$, as pointed out in \citen{Myers.R&Perry1986}. In fact, $D_1$ can be written as
\Eq{
D_1=\Delta(x+a^2\sin^2\theta+b^2\cos^2\theta+\mu ) + \mu ^2 x.
}
From this, around $x=x_h$ at which $\Delta=0$, the sign of $D_1$ is determined by the sign of $x_h$. Hence, when $x_h<0$, 2-tori generated by $\pd_{\phi_1}$ and $\pd_{\phi_2}$ become timelike surfaces around $x=x_h$, which always contain CTCs. This feature together with the fact that $\Phi$ has a finite non-degenerate limit $x=x_h$ implies that $x=x_h$ is not a horizon, as we show in \S\ref{sec:repulsons}.

\section{Disk Bundle over $\CP^{N-1}$}
\label{sec:DiskBundle}

In this section, we explain some basic features of the lens space type manifold  $S^{2N-1}/\ZR_n$ regarded as a $S^1$ bundle over $\CP^{N-1}$ and disk bundles over $\CP^{N-1}$ obtained from them. They are used to construct repulsions from a special class of Myers-Perry solutions in the next section.

\subsection{$S^1$ fibring of $S^{2N-1}$}

The unit sphere $S^{2N-1}$ can be embedded into $\CF^N$ by
\Eq{
\bm{z}\cdot\bm{\b z}\equiv \sum_{j=1}^N z_j\b z_j=1,
}
where $\bm{z}$ and $\bm{\b z}$ are the column vectors with $N$ entries, $\bm{z}=(z_j)$ and $\bm{\b z}=(\b z_j)$, respectively. If we parametrise $z_j$ as
\Eq{
z_j= \mu_j e^{i\phi_j},
}
this equation is expressed as
\Eq{
\sum_{j=1}^N \mu_j^2=1.
}
In terms of these coordinates the metric of $S^{2N-1}$ can be written
\Eq{
ds^2(S^{2N-1})= d\bm{z}\cdot d\bm{\b z}
  =\sum_{j=1}^N\inpare{ d\mu_j^2 + \mu_j^2d\phi_j^2}.
}

We can define a natural free $\UG(1)$ isometric action  on $S^{2N-1}$ by
\Eq{
z_j \mapsto e^{i\lambda } z_j.
}
As is well-known, the quotient space of $S^{2N-1}$ by this action is $\CP^{N-1}$, and the original $S^{2N-1}$ can be regarded as a $S^1$ bundle over $\CP^{N-1}$ with Fubini-Study metric. 

The explicit form of the metric in this fibring can be obtained in the following way. First, the infinitesimal transformation $X$ with unit norm corresponding this $\UG(1)$ action is given by
\Eq{
X=i \inpare{z_j \pd_j -\b z_j \b\pd_j},
}
Then, this vector field $X$ with a set of $\UG(1)$-invariant unit vector fields that are orthogonal to $X$ and project onto an orthonormal frame on $\CP^{N-1}$ form an orthonormal basis of vector fields on $S^{2N-1}$. Let $\chi^j$ be the 1-form basis dual to it such that $\chi^{2N-1}=\chi$ is dual to $X$. Then, $\chi$ is expressed as
\Eq{
\chi= \Im \bm{\b z}\cdot d\bm{z} = \sum_{j=1}^N \mu_j^2 d\phi_j,
\label{chi:def}
}
and the metric of $S^{2N-1}$ can be written
\Eq{
ds^2(S^{2N-1})= \chi^2 + ds^2(\CP^{N-1}),
\label{metric:S^{2N-1}:S1decomposition}
}
where $ds^2(\CP^{N-1})$ is the Fubini-Study metric of $\CP^{N-1}$ that is expressed in terms of the homogeneous coordinates $z_j$ as
\Eq{
ds^2(\CP^{N-1}) = \frac{d\bm{z}\cdot d\bm{\b z}}{\bm{z}\cdot \bm{\b z}}
  -\frac{|\bm{z}\cdot d\bm{\b z}|^2}{(\bm{z}\cdot \bm{\b z})^2}.
\label{metric:Fubini-Study:homogeneous}
}

Here, note that the K\"ahler form of this metric
\Eq{
\varphi = \frac{i}{2 \bm{z}\cdot \bm{\b z}} \sum_j dz_j \w d \b z_j
  -\frac{i}{2(\bm{z}\cdot \bm{\b z})^2}\bm{\b z}\cdot d\bm{z}\w \bm{z}\cdot d\bm{\b z}
\label{KahlerForm:CP^{N-1}}
}
is related to $\chi$ by
\Eq{
d\chi =2\varphi.
}
Further, it is easy to see that the set of isometries of $S^{2N-1}$ that preserves $\chi$ is isomorphic to $\UG(N)$, which projects onto the isometry group of $\CP^{N-1}$, $\SU(N)$\cite{Kobayashi.S&Nomizu1969B}. 

\subsection{$S^{2N-1}/\ZR_n$ and a disk bundle over $\CP^{N-1}$}
\label{subsec:DiskBundle}

Utilising the $S^1$ bundle structure of $S^{2N-1}$ introduced in the previous subsection, we can deform the standard metric of $S^{2N-1}$ to a deformed $\UG(N)$-symmetric one,
\Eq{
ds^2= c^2 \chi^2 + e^2 ds^2(\CP^{N-1}),
\label{metric:S1BundleOverCP}}
where $c$ and $e$ are non-vanishing constant. This is globally regular on $S^{2N-1}$ for any values of $c$ and $e$ if $ce\neq0$. 

Now, let us consider a disk bundle over $\CP^{N-1}$ obtained from this $S^1$ bundle with the metric
\Eq{
ds^2= d\xi^2 + c^2 \xi^2 \chi^2 + e^2 ds^2 (\CP^{N-1}).
\label{metric:DiskBundleOverCP}
}
It is clear that this disk bundle is regular except on the submanifold $\xi=0$. In order to see whether the metric is regular there or not, let us introduce the local charts $\xi_{(j)}^k$ ($k=1,\cdots, N, k\neq j$) and $w_{(j)}$ defined in terms of the coordinate $z_l=\mu_l e^{i\phi_k}$ by
\Eqrsubl{DiskBundle:chart}{
&& \xi_{(j)}^k = \frac{\mu_k}{\mu_j} e^{i(\phi_k-\phi_j)},\\
&& w_{(j)}= \xi e^{ic \phi_j},
}
where $j=1,\cdots,N$ is a label of the charts. The $j$-the chart cover the region $z_j\neq0$, and in terms of this local coordinates, the metric \eqref{metric:Fubini-Study:homogeneous} of $\CP^{N-1}$ can be written
\Eq{
ds^2(\CP^{N-1})= \frac{d \xi_{(j)}\cdot d {\b\xi}_{(j)}}{1+\xi_{(j)}\cdot\b\xi_{(j)}}
   -\frac{|\b\xi_{(j)} \cdot d\xi_{(j)}|^2}{(1+\xi_{(j)}\cdot\b\xi_{(j)})^2},
}
where $\xi_{(j)}=(\xi^k_{(j)})\in \CF^{N-1}$. Further, the disk part of the metric \eqref{metric:DiskBundleOverCP} can be written
\Eqr{
&& d\xi^2+c^2 \xi^2\chi^2
 =dw_{(j)}d\b w_{(j)}  -i c \inpare{\b w_{(j)} dw_{(j)}-w_{(j)}d\b w_{(j)}} \alpha 
  + |w_{(j)}|^2 \alpha^2,\\
&& \alpha = \frac{\b\xi_{(j)}\cdot d\xi_{(j)}-\xi_{(j)}\cdot d\b\xi_{(j)}}{2i\inpare{1+\b\xi_{(j)}\cdot\xi_{(j)}}}
}

Thus, the metric \eqref{metric:DiskBundleOverCP} is regular in each chart. 
In particular, because the submanifold $\xi=0$ is covered by these charts, the metric does not have a curvature singularity of any type at $\xi=0$. However, we can show that it has a quasi-regular singularity 
at $\xi=0$ for a generic value of $c$. 
In fact, if we look at the transformation between the $j_1$-th and the $j_2$-th coordinate systems,
\Eqrsubl{CoordTrf:DiskBundle}{
&& \xi_{(j_2)}^k = \frac{1}{\xi_{(j_1)}^{j_2}} \xi_{(j_1)}^k,\ (k\neq j_1,j_2),\\
&& \xi_{(j_2)}^{j_1} = \frac{1}{\xi_{(j_1)}^{j_2}},\\
&& w_{(j_2)}= w_{(j_1)}\pfrac{ \xi_{(j_1)}^{j_2}}{|\xi_{(j_1)}^{j_2}|}^{-c},
}
we find that the coordinate transformation becomes regular and well-defined only when $c$ is an integer. We can avoid this problem and construct a consistent set of charts around $\xi=0$ if we use some non-zero integer instead of $c$ in \eqref{DiskBundle:chart}, but in this case, the expression for the metric \eqref{metric:DiskBundleOverCP} in each chart becomes singular at $w_{(j)}=0$ for a non-integer value of $c$ because the singular part of $d\xi^2=\inpare{\Re (dw_{(j)}/w_{(j)})}^2$ is not cancelled by  $c^2\xi^2 \chi^2$. 

To see what is happening, first note that the coordinates $\xi_{(j)}$ and $w_{(j)}$ are invariant under the transformation
\Eq{
\phi_j \mapsto \phi_j + \frac{2\pi}{c},\ (j=1,\cdots,N)
}
for $c\neq0$. However, the transformation group generated from this transformation becomes properly discontinuous on each $S^{2N-1}$ corresponding to a fixed value of $\xi\neq0$ only when $|c|$ is a positive integer $n$. Further, in that case, the manifold reconstructed from the charts is a disk bundle corresponding to $S^{2N-1}/\ZR_n$. If we introduce the angle coordinate $\psi$ by
\Eq{
\psi=\frac{1}{N}\sum_{j=1}^N \phi_j,
}
$\chi$ can be written as
\Eq{
\chi = d\psi + \beta,
}
where $\beta$ is a 1-form on $\CP^{N-1}$. The above transformation acts on $\psi$ as $\psi\mapsto \psi+2\pi/n$ for $c=n$ and reduces the effective range of $\psi$ to $2\pi/n$. This is concordant to the fact that the regularity of the disk metric $d\xi^2 + c^2 \xi^2d\psi^2$ requires the $2\pi$ periodicity of the effective angle variable $c\psi$. In contrast, for a non-integer value of $c$, because the periodicity of $\psi$ is restricted to $2\pi/n$ for some integer $n$, the origin $\xi$ becomes a conical singularity.

In general,  $S^1$ bundles over $\CP^{N-1}$ are classified by the first Chern number defined by the integration of 2-form $F/(2\pi i)$, where $F$ is the curvature form  of the principal $\UG(1)$ bundle over $S^2\cong \CP^1\subset\CP^{N-1}$ corresponding to the unique generating element of $H_2(\CP^{N-1})$. Because $n\chi$ provides the connection form for the $S^1$ bundle over $\CP^{N-1}$ corresponding to $S^{2N-1}/\ZR_n$, its first Chern number is given by
\Eq{
\int_{\CP^1} \frac{1}{2\pi i} n d\varphi = \int_{\CP^1} \frac{n}{\pi i}\varphi = n.
}
Thus, the integer value for $c$ corresponds to the topological number.

\section{Repulsons}
\label{sec:repulsons}

In this section, we show that the special Myers-Perry solution with all momentum parameters equal has a  similar internal structure to the five-dimensional one discussed in \S\ref{sec:5DMP} for any odd spacetime dimensions $D\ge5$ and construct asymptotically flat regular soliton spacetimes with negative mass from the special Myers-Perry solution with negative mass. 

\subsection{$\UG(N)$ MP solution}

For an odd spacetime dimension $D=2N+1$, the Myers-Perry solution\cite{Myers.R&Perry1986} can be written
\Eq{
ds^2 = \frac{r^2}{\Delta} dr^2 + (r^2+a^2) ds^2(S^{2N-1}) + \frac{\mu}{(r^2+a^2)^{N-1}} (dt- a \chi)^2 - dt^2
\label{metric:U(N)MP}
}
where
\Eq{
\Delta = r^2+a^2 -\frac{\mu r^2}{(r^2+a^2)^{N-1}},
}
and $\chi$ is the $\UG(N)$-invariant 1-form \eqref{chi:def} on $S^{2N-1}$ introduced in \S\ref{sec:DiskBundle}. Note that the metric \eqref{metric:U(N)MP} has $\UG(N)$ invariance in addition to the time translation invariance.

In terms of the coordinate  $y$ defined by
\Eq{
y= r^2+a^2
}
this metric can be regularly extended to $r^2<0$ as\cite{Myers.R&Perry1986}
\Eqr{
ds^2 &=& \frac{d y^2}{4 \Delta}+ ds_K^2, \label{metric:U(N)MP:y}\\
ds_K^2 &=& 
  y ds^2(S^{2N-1}) + \frac{\mu}{ y^{N-1}} (d t- a \chi)^2 -  d t^2.
}
where $\Delta$ now reads
\Eq{
\Delta =  y-\frac{\mu ( y- a^2)}{ y^{N-1}}.
\label{Delta:def:N}
}

Utilising the decomposition \eqref{metric:S^{2N-1}:S1decomposition}, the metric of the Killing orbits $ds_K^2$ can be written
\Eq{
ds_K^2 = - C dt^2 + B (\chi -\Omega dt)^2+ y ds^2(\CP^{N-1}),
\label{metric:Killing:N}
}
where
\Eqrsubl{metric:Killing:N:coeff}{
B &=& \frac{ y^N+ \mu a^2  }{ y^{N-1}},\\
C &=& \frac{y^{N-1} \Delta}{y^N+ \mu a^2 },\\
\Omega &=& \frac{\mu a}{y^N + \mu a^2}.
}
From this, it follows that
\Eqrsub{
&& \det g_K= -y^{2N-2} \Delta \prod_j \mu_j^2,\\
&& \det g= -\frac{1}{4}y^{2N-2}\prod_j \mu_j^2.
}
Hence the possible loci of singularity are $y=0$ and $\Delta=0$. Among these, $y=0$ is a curvature singularity because the Kretchman invariant of this metric is given by
\Eq{
R_{\mu\nu\lambda\sigma}R^{\mu\nu\lambda\sigma}
 = \frac{4N(N-1)\mu^2}{ y ^{2N}}\inrbra{(2N-1)^2-8N(N+1)\frac{a^2}{ y}
   +4(N+1)(N+2)\frac{ a^4}{ y^2}}.
}
%

\begin{figure}
\centerline{
\includegraphics*[height=8cm]{\FigDir/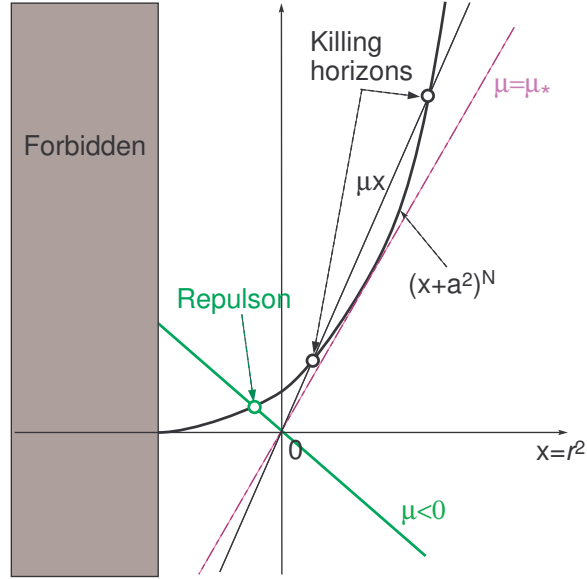}
}
\caption{Horizon position of the $\UG(N)$ MP solution}
\label{fig:Delta_U(N)}
\end{figure}

\subsection{Internal Structure}

As in the five-dimensional case, the location of the horizon is determined by the zero of $\Delta$,
\Eq{
(x+a^2)^{N-1}\Delta=(x+a^2)^N -\mu x=0
}
where $x=r^2$.  As is shown in Fig. \ref{fig:Delta_U(N)}, this equation has two positive roots for $\mu>\mu_*=N^N a^{2(N-1)}/(N-1)^{N-1}$, while it has a single negative root $x_h$ for $\mu<0$. In the the former case, there exists no CTC in the regular region $y=x+a^2>0$, and the two roots correspond to Killing horizons. 

In contrast, in the negative mass case, CTC appears in the region $\Delta>0$ from \eqref{Delta:def:N} and \eqref{metric:Killing:N}, and $g_{tt}>0$ around $x=x_h$. Hence, $x=x_h$ cannot be a horizon as in the $N=2$ case. Now, we show that it is a quasi-regular singularity in general that can be made regular by some periodic identification of the time coordinate for a discrete set of values for the angular momentum $a$. 

For that purpose, we introduce the new coordinate $\xi$ by
\Eq{
\xi= \int_{y_h} \frac{dy}{2\sqrt{\Delta}}.
\label{xi:def}}
%
At around $\xi=0$, $y$ is expressed in terms of $\xi$ as
\Eq{
y-y_h= \Delta'(y_h) \xi^2 + \Order{\xi^4}.
}
Hence, the metric \eqref{metric:U(N)MP:y} can be expanded around $\xi$ as
\Eqr{
ds^2
 &=& d\xi^2 + c^2\xi^2 \t\chi^2 + y ds^2(\CP^{N-1}) -\frac{y_h^2}{a^2-y_h}\inpare{1+\Order{\xi^2}}\t\tau^2 + \Order{\xi^4} 
 \notag\\
 &\approx & d\xi^2 + c^2\xi^2 \t\chi^2 + y_h ds^2(\CP^{N-1}) -\frac{y_h^2}{a^2-y_h}\t\tau^2,
\label{metric:U(N)MP:repulson}
}
where $c$ is the constant
\Eq{
c =\frac{aN}{\sqrt{a^2-y_h}}\inpare{1-\frac{N-1}{N}\frac{y_h}{a^2}},
}
and $\t\tau$ and $\t\chi$ are the 1-forms 
\Eqrsub{
\t\tau &=& \chi -\frac{a}{y_h} dt,\\
\t\chi &=& \chi + \frac{(N-1)y_h}{Na^2-(N-1)y_h} \t\tau.
}

\subsection{Construction of regular solitons}
\label{subsec:soliton}

As you immediately notice, the metric \eqref{metric:U(N)MP:repulson} is identical to the metric \eqref{metric:DiskBundleOverCP} with $e^2=y_h$ in directions orthogonal to $\t\tau$. Hence, it is expected that $\xi=0$ is a quasi-regular singularity in general and can be made regular when $c$ is an integer. We can confirm this expectation by extending the charts \eqref{DiskBundle:chart} for the disk bundle in \S\ref{subsec:DiskBundle} to include the time coordinate as
\Eqrsubl{Repulson:chart}{
&& \xi_{(j)}^k = \frac{\mu_k}{\mu_j} e^{i(\phi_k-\phi_j)},\\
&& w_{(j)}= \xi e^{ic \phi_j},\\
&& t_{(j)}= t-\frac{y_h}{a} \phi_j.
}
From the expression for $\t\tau$,
\Eq{
\t\tau= -\frac{a}{y_h} dt_{(j)}
 +\frac{\b\xi_{(j)}\cdot\xi_{(j)}-\xi_{(j)}\cdot d\b\xi_{(j)}}{1+\b\xi_{(j)}\cdot\xi_{(j)}},
}
and the argument in \S\ref{subsec:DiskBundle}, it is easy to see that the metric \eqref{metric:U(N)MP:repulson} can be expressed regularly in terms of this set of coordinates for any $(j)$. Hence, the locus $x=x_h$ is at most a quasi-regular singularity, and its regularity is equivalent to the regularity of the coordinate transformations. For the pair of charts $(j_1)$ and $(j_2)$, it is given by
\Eqrsubl{CoordTrf:Repulson}{
&& \xi_{(j_2)}^k = \frac{1}{\xi_{(j_1)}^{j_2}} \xi_{(j_1)}^k,\ (k\neq j_1,j_2),\\
&& \xi_{(j_2)}^{j_1} = \frac{1}{\xi_{(j_1)}^{j_2}},\\
&& w_{(j_2)}= w_{(j_1)}\pfrac{ \xi_{(j_1)}^{j_2}}{|\xi_{(j_1)}^{j_2}|}^{-c},\\
&& t_{(j_2)}= t_{(j_1)} + i\frac{y_h}{a} \ln \pfrac{z_{(j_1)}^{j_2}}{|z_{(j_1)}^{j_2}|}.
}
As in the disk bundle case, from the well-definedness of the transformation for $w_{(j)}$, $c$ is constrained to be an integer:
\Eq{
\frac{aN}{\sqrt{a^2-y_h}}\inpare{1-\frac{N-1}{N}\frac{y_h}{a^2}}=n.
}
This condition can be written
\Eq{
\pfrac{y_h}{a^2}^2 -\frac{2N(N-1)-n^2}{(N-1)^2}\frac{y_h}{a^2}
 + \frac{N^2-n^2}{(N-1)^2}=0.
\label{eq:yh/a^2}}
For each solution to this, the mass and the angular parameters are related as
\Eq{
\frac{a^{2(N-1)}}{|\mu|}=\frac{1-y_h/a^2}{(y_h/a^2)^N}.
}
For example, for $N=2$ or $N=3$, i.e. for $D=5$ or $D=6$, the equation \eqref{eq:yh/a^2} has a positive root only for $n=3,\cdots$ and it is unique. For $N\ge4$, it has two positive roots for $2\sqrt{N-1}\le n<N$ and a single positive root for $n\ge N$. 

This condition on $c$ is not sufficient for the regularity, because the time coordinate transformation in \eqref{CoordTrf:Repulson} has an additive ambiguity due to the phase term. This ambiguity can be made harmless only when we introduce the periodic identification of each time coordinate $t_{(j)}$ by
\Eq{
t_{(j)}\sim t_{(j)}+ \frac{2\pi y_h}{m a},
}
where $m$ is a positive integer. By this identification, $\exp(im a t_{(j)}/y_h)$ becomes a good time coordinate and the coordinate transformations can be expressed as
\Eq{
e^{im at_{(j_2)}/y_h}= e^{im at_{(j_1)}/y_h}\pfrac{z_{(j_1)}^{j_2}}{|z_{(j_1)}^{j_2}|}^{-m}.
}
Hence, each $\xi=\const$ slice of the new regular spacetime obtained by this identification is a torus body bundle over $\CP^{N-1}$. Because the disk bundle is simply connected and the $S^1$ bundle in the $\t\tau$-direction over $\CP^{N-1}$ is homeomorphic to $S^{2N-1}/\ZR_m$, the fundamental group of the whole spacetime is isomorphic to $\ZR_m$. In particular, for the choice $m=1$, this spacetime is simply connected, regular and asymptotically flat, although its spatial infinity is homeomorphic to $S^1\times S^{2N-1}/\ZR_n$. 

In this section, we only proved the regularity of the repulson spacetime in the local sense. Generally speaking, this may not guarantee the regularity in a global sense, such as the geodesic completeness. In Appendix \ref{app:geodesics}, we prove that our repulson spacetimes are geodesically complete for any geodesics.

\section{Summary and Discussion}

In the present paper, we have shown that the Myers-Perry solution with negative mass describes an asymptotically flat rotating spacetime with naked CTCs and no curvature singularity. We further pointed out that we can construct a regular repulson-type asymptotically flat soliton from them for a discrete set of values of the angular momentum for a fixed mass parameter $\mu$. 

  Although the structure of the original metric when the repulson appears is similar to that of BPS solutions with repulsons, there exist several big differences between our family and the BPS families. Firstly, our solutions satisfy the purely vacuum Einstein equations. This is an important point because the existence of matter, especially a gauge field tend to help the appearance of CTCs as the four dimensional solutions mentioned above indicate. Secondly, our solutions do not have any supersymmetry and the repulson becomes a regular submanifold of the whole spacetime. This should be contrasted to the BPS case in which the repulson is at spatially infinite distance. Thirdly, due to this feature, we had to introduce periodic identifications both in an angular direction and in the time direction in order to remove the quasi-regular singularity at the repulson. As a consequence, the whole spacetime obtained after this regularisation is not asymptotically flat in the standard sense because it has some time periodicity at spatial infinity and its spatial infinity is $S^{2N-1}/\ZR_n$ with $n\ge2$, although it is topologically simply connected. Fourthly, the repulson appears only for negative mass. This is rather striking because a regular soliton spacetime with negative mass can exist in higher dimensions. This does not contradict the positive energy theorem because the spacetime has causality violation as discussed in \citen{Myers.R&Perry1986} and moreover is not globally asymptotically flat. Finally, the quasi-regular singularity can be removed only for a set of discrete values. 

  In general, the existence of a regular asymptotically flat solution with negative mass produces instability and is very dangerous. In the case of our solutions, however, since it has a non-trivial topological asymptotic structure, it is not so obvious whether they might provoke some kind of instability classically or quantum mechanically.

  Finally, we comment on the extension of our analysis. In the present paper, we have explicitly studied only the case in which all the angular momentum parameters are equal for simplicity, but this restriction does not appear to be essential. It will be also interesting to apply similar construction to the rotating black-hole type solutions with cosmological constant\cite{Gibbons.G&&2005}. Because the cosmological constant does not have an essential influence on the internal structure of the black hole solutions, there will arise no essential new problem in such construction, although some new features may appear concerning the global structure.

\section*{Acknowledgement}

The authors  would like to thank Hideki Maeda, Akihiro Ishibashi, Shinya Tomizawa and Andres Anabalon for discussions and useful comments. HK was supported in part by Grants-in-Aids for Scientific Research from JSPS (No. 18540265) and for the Japan-U.K. Research Cooperative Program.

\appendix

\section{Extension across Horizon}
\label{app:horizon}

We can show that when \eqref{horizoneq:x} has a positive root, the surface $x=x_h$ corresponding to the bigger real root is actually a horizon of the solution \eqref{metric:5DMP:BL} for $\mu>(|a|+|b|)^2$. The most direct method is to introduce the following new coordinates $u,\phi_{1+},\phi_{2+}$:
\Eqrsub{
dt &=& du -\frac{(x+a^2)(x+b^2)}{\Delta} \frac{dx}{2r_h},\\
d{\phi_1} &=& d\phi_{1+} + \frac{a(x+b^2)}{\Delta}\frac{dx}{2r_h}
 - \frac{a du}{r_h^2+a^2},\\
d{\phi_2} &=& d\phi_{2+} + \frac{b(x+a^2)}{\Delta}\frac{dx}{2r_h}
 - \frac{b du}{r_h^2+b^2}.
}
Then, the metric can be expressed as
\Eqr{
ds^2 &=& -\frac{\rho dx^2}{4(x_h-a^2b^2)}
  +\inrbra{\frac{x_h \rho_h du}{(x_h+a^2)(x_h+b^2)}
   + a\sin^2\theta d\phi_{1+}  + b\cos^2\theta d\phi_{2+}}
   \frac{dx}{r_h}
   \notag\\
 && + \frac{(x-x_h)}{(x_h+a^2)(x_h+b^2)\rho^2}
 \Big\{-\frac{P du}{(x_h+a^2)(x_h+b^2)}
 \notag\\
 &&
   -2a(x_h+b^2)(\rho^2+a^2+x_h)\sin^2\theta d\phi_{1+}
   \notag\\
 &&
   -2b(x_h+a^2)(\rho^2+b^2+x_h)\cos^2\theta d\phi_{2+}
   \Big\} du 
   \notag\\
 &&
   +\rho^2d\theta^2
   \notag\\
 && +\Big\{ x^2+ \inpare{a^2+a^2\cos^2\theta+b^2\sin^2\theta}x
  + a^4 + 2a^2b^2 \sin^2\theta 
  \notag\\
 && 
   +\frac{a^2}{x_h}(x_h^2+a^2b^2)\sin^2\theta \Big\}
   \frac{\sin^2\theta}{\rho^2}d\phi_{1+}^2
   \notag\\
 && +\Big\{ x^2+ \inpare{b^2+b^2\sin^2\theta+a^2\cos^2\theta}x
  + b^4 + 2a^2b^2 \cos^2\theta 
  \notag\\
 && 
   +\frac{b^2}{x_h}(x_h^2+a^2b^2)\cos^2\theta\Big\}
   \frac{\cos^2\theta}{\rho^2}d\phi_{2+}^2
   \notag\\
 && + \frac{2ab(x_h+a^2)(x_h+b^2)\sin^2\theta\cos^2\theta}{x_h}
 \frac{d\phi_{1+} d\phi_{2+}}{\rho^2}.
}
Here
\Eqr{
P &=& (a^2-b^2)(a^2b^2-x_h^2)\inpare{\cos^2\theta
 + 2b^2 + x_h + x}\cos^2\theta
 \notag\\
 && + (x_h+b^2)^2 \inpare{a^2b^2-a^2 x_h + a^2 x -x_h^4}.
}
%

\section{Geodesic Completeness of the Repulson Spacetime}
\label{app:geodesics}

In this appendix, we prove the geodesic completeness of the regular soliton spacetime constructed in \S\ref{sec:repulsons}.

\subsection{Geodesic equations}

For the metric \eqref{metric:U(N)MP:y} with \eqref{metric:Killing:N}, the geodesic equations can be derived as the equation of motion for the Lagrangian
\Eq{
\LL = \frac{\d y^2}{8\Delta} - \frac{C}{2}\d t^2 + \frac{B}{2}(\d\chi-\Omega \d t)^2
 +\frac{y}{2} \d\sigma^2,
}
where $B$ and $C$ are functions of $y$ defined in \eqref{metric:Killing:N:coeff}, the dot implies the differentiation with respect to an affine parameter $\lambda$, $\d\chi$ denotes the value of $\chi$ at the tangent vector of the geodesic, and $\d\sigma^2$ is defined in terms of the metric of $\CP^{N-1}$ and its affine coordinates $u^a$ as
\Eq{
\d\sigma^2= \gamma_{ab}(u)\d u^a \d u^b.
}
In one of the charts introduced in \S\ref{subsec:soliton}, $\d\chi$ can be written in the form
\Eq{
\d\chi= \d\phi + \d\beta=\d\phi + \beta_{a}\d u^a,
}
where $\phi$ is some of the angle variable $\phi_j$, and $\beta$ is a 1-from on $\CP^{N-1}$. 

The momenta conjugate to $t$, $\phi$, $u^a$ and $y$ are
\Eqrsub{
&& E= C\d t + L \Omega,\\
&& L= B(\d\chi-\Omega \d t),\\
&& p_a = L\beta_a + y \gamma_{ab}\d u^b,\\
&& p_y = \frac{\d y}{4\Delta}.
}
Among these, because $t$ and $\phi$ are cyclic coordinates, $E$ and $L$ are conserved. The other momenta obey the equations
\Eqrsub{
&& \d p_a = L \pd_a \beta_b \d z^b + \frac{y}{2}\pd_a\gamma_{bc}\d z^b \d z^c,\\
&& \d p_y = -\frac{\Delta'}{8\Delta^2} \d y^2 - \frac{C'}{2}\d t^2 + \frac{B'}{2B^2} L^2 
   - L\Omega' \d t + \frac{1}{2} \d \sigma^2.
}
In terms of the other equation of motion and the conservation laws, the first of these can be rewritten 
\Eq{
\d q_a = 2L \varphi_{ab}\d z^b + \frac{y}{2} \pd_a \gamma_{bc}\d z^b \d z^c,
}
where
\Eq{
q_a= y \gamma_{ab}\d z^b,
}
and $\varphi_{ab}$ is the K\"ahler 2-form \eqref{KahlerForm:CP^{N-1}}. From this, it follows that 
\Eq{
J^2= q_a q_b \gamma^{ab}= y^2\d\sigma^2
}
is conserved. This conservation law is a consequence of the $\UG(N)$ symmetry of the Fubini-Study metric of $\CP^{N-1}$. 

Utilising the conservation equations including this and the equation for $p_y$ or the affine parameter condition, we obtain the following 1st-order equation for $\xi$ defined by \eqref{xi:def}:
\Eq{
\frac{1}{2}\d\xi^2 + V(\xi)=\epsilon (=0,\pm1),
\label{1stInt:xi}
}
with
\Eq{
V=\frac{1}{\Delta}\insbra{L^2 - E^2 y -\frac{\mu(aE-L)^2}{y^{N-1}}} + \frac{J^2}{y},
\label{V:def}
}
where $y$ is regarded as a function of $\xi$. For each solution $\xi(\lambda)$ of this equation, the behavior of the other coordinates along the geodesic is determined by integrating the equations
\Eqrsub{
&& \d t = \frac{E-L\Omega}{C}, \label{geodesiceq:t}\\
&& \d u^a= \frac{1}{y} \gamma^{ab} q_b,\label{geodesiceq:u}\\
&& D_{\d u} q_a= \frac{2L}{y} \psi_{ab}q^b,\label{geodesiceq:q}\\
&& \d\phi = -\beta_a\d u^a + W, \label{geodesiceq:phi}
}
where 
\Eq{
W=\frac{\mu(aE-L)+ L y^{N-1}}{y^{N-1} \Delta}.
}
%

\subsection{Behaviour of geodesics}

We start from the analysis of \eqref{1stInt:xi}. It is clear that the numerator of the 1st term of the potential \eqref{V:def} is monotonically decreasing function unless $E=L=0$, for which it identically vanishes. Further, its value at $y=y_h$ is given by
\Eq{
\frac{(La-Ey_h)^2}{a^2-y_h}.
}
Hence, except for the case $La=E y_h$, the effective potential $V$ becomes positive infinity as $y$ approaches $y_h$. This implies that only geodesics whose constants of motion satisfying the relation 
\Eq{
La=E y_h
\label{nonrepulsive}
}
can reach the submanifold $y=y_h$. 

The other geodesics are reflected at some point $y>y_h$ and go away to infinity or oscillate in some region that does not contain the repulson submanifold. Since the corresponding solution can be obtain just by the integration
\Eq{
\lambda= \int \frac{d\xi}{\sqrt{\epsilon - V}}
   =\int\frac{dy}{\sqrt{2\Delta(\epsilon-V)}},
\label{geodesic:sol:y}}
and $V$ is a meromorphic function of $y$, the solution can be always infinitely extendible. 

Hence, only a geodesic approaching the repulson submanifold may be incomplete. Equation \eqref{geodesic:sol:y} still applies to such a geodesic, and the incompleteness may occur only when $\epsilon-V$ vanishes at $y=y_h$ linearly in $\xi$. This cannot happen because $V$ is a meromorphic function of $y$ and $y-y_h$ is quadratic in $\xi$ in the leading order. Therefore, all geodesics go through $y=y_h$ at a finite affine time and pass away. 

We also need to check that the other coordinates of each geodesic do not behave pathologically. First, from \eqref{geodesiceq:u} and \eqref{geodesiceq:q}, we find that no pathology occurs in the $\CP^{N-1}$ sector. Further, from \eqref{geodesiceq:phi} and the regularity of $W$ at $y=y_h$ in the case \eqref{nonrepulsive}, 
\Eq{
W(y_h)= \frac{E(N-1)(a^2-y_h)}{a\inrbra{y_h + N(a^2-y_h)}},
}
no pathology occurs for $\phi$ as well. Similarly, in the equation for $t$, \eqref{geodesiceq:t}, we can show that the right-hand side of this equation is regular and finite except at $y=y_h$, which becomes also regular under the condition \eqref{nonrepulsive}. Thus, we can conclude that all geodesics in the regular repulson spacetimes constructed in \S\ref{sec:repulsons} are complete.


\end{document}